\documentclass{article}

\usepackage{preprint}
\usepackage[utf8]{inputenc} 
\usepackage[T1]{fontenc}    
\usepackage{hyperref}       
\usepackage{url}            
\usepackage{booktabs}       
\usepackage{amsfonts}       
\usepackage{nicefrac}       
\usepackage{microtype}      
\usepackage{fancyhdr}       
\usepackage{graphicx}       
\usepackage{placeins}
\usepackage[backend=biber, sorting=none]{biblatex}
\usepackage{caption}

\graphicspath{{./Figures/}}     
\newcommand{\beginsupplement}{%
        \setcounter{table}{0}
        \renewcommand{\thetable}{S\arabic{table}}%
        \setcounter{figure}{0}
        \renewcommand{\thefigure}{S\arabic{figure}}%
     }

\title{Brain-inspired polymer dendrite networks for morphology-dependent computing hardware}

\author{
  Corentin Scholaert\textsuperscript{a}, Yannick Coffinier\textsuperscript{a}, Sébastien Pecqueur\textsuperscript{a}, Fabien Alibart\textsuperscript{a,b}  \\
  \\
  a. IEMN, UMR 8520 -   Univ. Lille, CNRS, Univ. Polytechnique Hauts-de-France -   59000 Lille, France\\
  \\
  b. Laboratoire Nanotechnologies \& Nanosystèmes (LN2) -  CNRS IRL-3463, 3IT - Québec J1K0A5, Sherbrooke, Canada\\
  \\
  \texttt{fabien.alibart@univ-lille.fr} \\
}

\makeatletter
\AtBeginDocument{%
	\expandafter\renewcommand\expandafter\subsection\expandafter
	{\expandafter\@fb@secFB\subsection}%
	\newcommand\@fb@secFB{\FloatBarrier
		\gdef\@fb@afterHHook{\@fb@topbarrier \gdef\@fb@afterHHook{}}}%
	\g@addto@macro\@afterheading{\@fb@afterHHook}%
	\gdef\@fb@afterHHook{}%
}
\makeatother

\begin{document}
\maketitle

\begin{abstract}
Variability has always been a challenge to mitigate in electronics. This especially holds true for organic semiconductors, where reproducibility and long-term stability concerns hinder industrialization. By relying on a bio-inspired computing paradigm, we show that AC-electropolymerization is a powerful platform for the development of morphology-dependent computing hardware. Our findings reveal that electropolymerized polymer dendrite networks exhibit a complex relationship between structure and operation that allows them to implement nearly linear to nonlinear functions depending on the complexity of their structure. Moreover, dendritic networks can integrate a limitless number of inputs from their environment, for which their unique morphologies induce specific patterns in the dynamic encoding of the network’s output. We demonstrate that this property can be used to our advantage in the context of \textit{in materio} computing to discriminate between different spatiotemporal inputs. These results show how, due to its inherent stochasticity, electropolymerization is a pivotal technique for the bottom-up implementation of computationally powerful objects. We anticipate this study will help shifting the negative perception of variability in the material science community and promote the electropolymerization framework as a foundation for the development of a new generation of hardware defined by its topological richness.
\end{abstract}

\section{Introduction}
Silicon-based digital computing technologies have achieved tremendous levels of sophistication and perfection over a relatively short period of time in the realm of digital information processing. Yet, digital computing is only a small corner of the much larger field of general computing. For some time now, researchers have proposed unconventional ways to perform computing, redefining the notion as establishing input-output relationships and tapping into the intrinsic physics of virtually any physical system to perform computing tasks\cite{tero_rules_2010, qi_physical_2023, usami_materio_2021, adamatzky_physarum_2007, finocchio_roadmap_2024}. Within the general framework of unconventional computing, brain-inspired approaches have lately started to attract a lot of attention\cite{ham_neuromorphic_2021, schuman_opportunities_2022, sangwan_neuromorphic_2020, zhu_comprehensive_2020, aimone_review_2022}. Neuromorphic computing could offer some answers to the challenges faced by digital computing as the brain is able to massively parallelize information processing at a very low energy cost, all the while being robust and adaptable thanks to its structural and functional plasticity.   
Recently, a new generation of neuromorphic devices based on organic electronic materials has been developed, drawing inspiration from and mimicking the operation of the brain\cite{gkoupidenis_neuromorphic_2015, rondelli_pre-synaptic_2023, van_de_burgt_organic_2018, krauhausen_organic_2021, gerasimov_evolvable_2019, gkoupidenis_organic_2022, matrone_modular_2024, van_doremaele_retrainable_2023,gerasimov_biologically_2023}. These devices hold the potential to reach ultralow energy consumption\cite{duan_low-power-consumption_2020, lee_organic_2021} and are easier to process than traditional silicon-based technology\cite{donahue_tailoring_2020}. When it comes to fabrication, one of the most interesting aspects of organic electronics is that monomers can simply be polymerized, which renders the fabrication process much easier. Indeed, over the past few years, electropolymerization has proven an increasingly promising fabrication technique, allowing for a more adaptative bottom-up approach. It opens the way for the easy and versatile deposition of conductive polymer coatings\cite{koizumi_electropolymerization_2016, watanabe_-plane_2018, koizumi_synthesis_2018, janzakova_analog_2021, chen_ac-bipolar_2023}, which have been used to implement various neuromorphic features, such as sensory devices\cite{ghazal_electropolymerization_2023, ghazal_precision_2023}, plastic synapses\cite{gerasimov_evolvable_2019, gerasimov_biologically_2023} and spiking neurons\cite{harikesh_organic_2022}. Yet, in these cases, organic electrochemistry is only perceived an alternative tool for the deposition of polymer thin films.\\

One of the main drawbacks of organic electronics has always been its instability and variability, in part due to the phase changes associated with the low transition temperatures of soft mater. Although low-cost, the commercialization of organic electronic materials is still minimal because of operational and  long-term ambient stability concerns\cite{chen_ambient_2021, lee_toward_2017}. Yet, this inherent variability of soft mater could turn beneficial. Electropolymerization on organic electrochemical transistors (OECTs) has been exploited to generate structurally different active materials on lithographically patterned devices in order to enrich the dynamics of a reservoir, thus facilitating the classification of dynamic information\cite{pecqueur_neuromorphic_2018}. This was possible because each OECT of the array was unique, therefore projecting the information on a wider dimensional space. More recently, networks of electropolymerized polymer fibers have been shown to implement Hebbian learning\cite{hagiwara_fabrication_2023} and Boolean logic\cite{akai-kasaya_evolving_2020}, as well as biosignal classification through the reservoir computing framework\cite{cucchi_reservoir_2021}. These systems therefore appear to be interesting candidates for the implementation of unconventional computing frameworks, capitalizing on the complex dynamics at work within these networks of artificial dendrites.\\

However, to this day, organic devices still do not exploit the topological diversity offered by the electropolymerization framework. The rich structural variety offered by this bottom-up approach could be taken advantage of to perform \textit{in materio} computing based on the unique behavior of each device. Therefore, electropolymerization is not only a cost-effective alternative to otherwise expensive fabrication techniques, but also a whole new avenue for the conception of a class of a neuromorphic hardware with morphology-dependent functions.\\

In this article, we explore the possibilities offered by poly(3,4-ethylenedioxythiophene):polystyrene sulfonate (PEDOT:PSS) fiber networks from an unconventional computing point of view by physically implementing hardware-dependent functions with a unique structure-operation relationship. We characterize and exploit the topological richness of electrogenerated conducting polymer dendrites for \textit{in liquido} computation. In particular, electrochemical self-gating of single dendrites and inter-gating effects cause voltage-to-current nonlinear transformations that can be used to our advantage to process information. We show that these systems present an inherent form of memory that can be electrochemically programmed by specifically organizing the ion space charges in the polymer network. Finally, we show that such networks can be used as an electrochemical classifier, where spatiotemporal projections of nonlinearly separable input patterns can be discriminated thanks to the higher dimensional projection inherent to the dendritic network complexity and exploiting \textit{in materio} computational resources. These properties make such dendritic networks advantageous for the development of hardware for artificial intelligence such as the physical implementation of reservoirs, and are also sought-after elements for the realization of Physical Unclonable Functions (PUF) with embedded computational functionalities\cite{bohm_physical_2013}.

\section{Results}

\subsection{Nonlinear behavior of physically connected dendrites}
Previous studies had brought to light that conducting polymer dendrites behave like OECTs\cite{janzakova_dendritic_2021}: applying a positive voltage to the gate electrode of the system leads cations into the bulk of the material, thus dedoping the PEDOT:PSS channel. So far, these studies were mostly limited to one electrochemical transistor, with a continuous dendritic channel grown between two electrodes. Cucchi and coworkers went a step further and took advantage of the nonlinearities observed in networks of polymer fibers to implement reservoir computing\cite{cucchi_reservoir_2021}. In their implementation, the authors took advantage of feedback connections to increase the nonlinearity of the spatiotemporal signal projection, thus requiring additional circuitry overhead. In this article, we present a solution where nonlinearities are induced uniquely by the complexity of network topologies, and the resulting auto-gating and inter-gating effects mediated through the common electrolyte.\\
 
\begin{figure}[h]
  \centering
  \includegraphics[width=1\columnwidth]{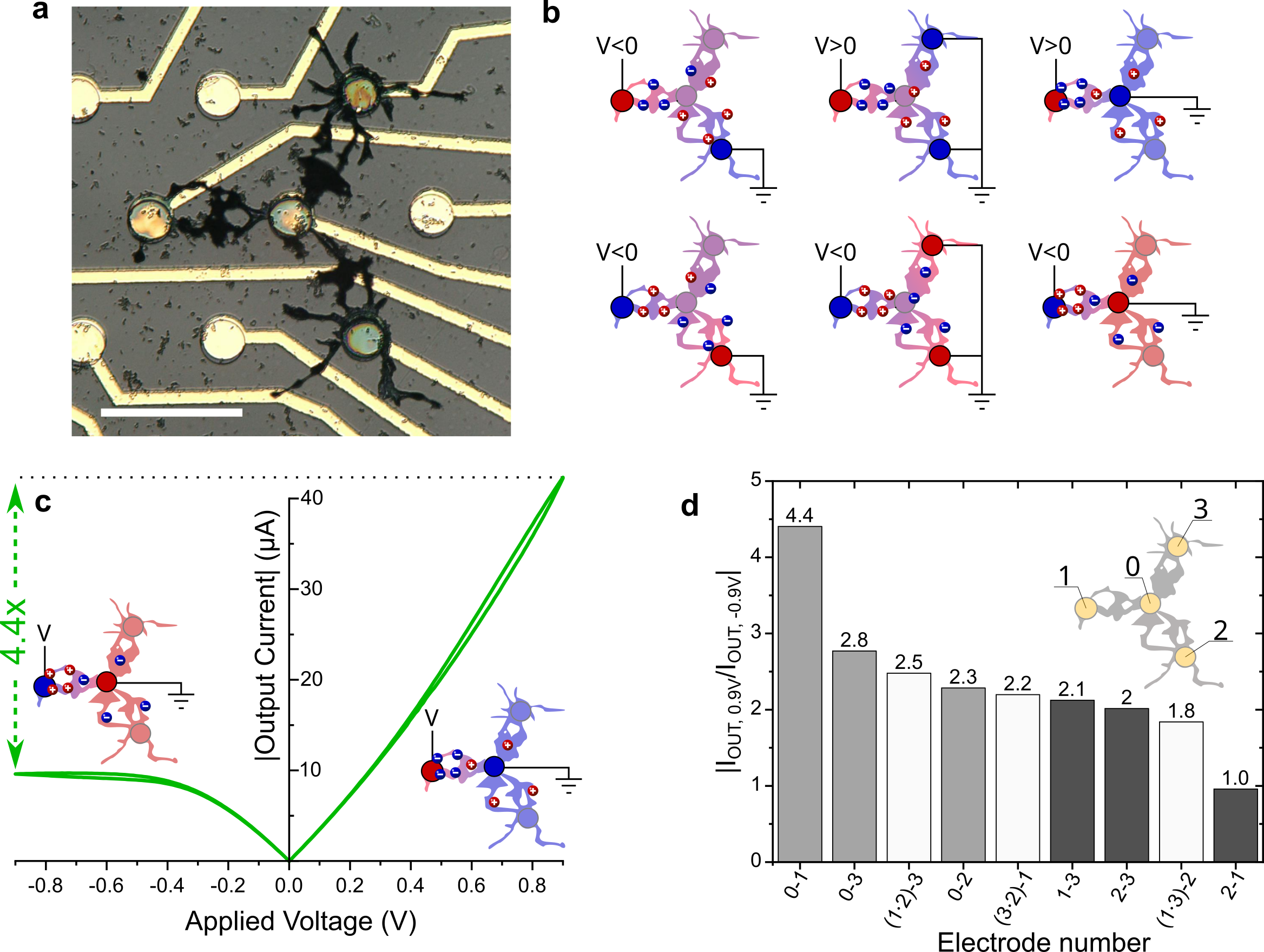}
  \caption{\textbf{Nonlinear behavior and autogating effect of a multiterminal dendritic OECT. a,} Microscopic photo of a four-terminal dendritic system. Scale bar = 100 µm. \textbf{b,} Schematic of ionic distribution within the system when different potentials are applied. Relative potential is represented by the color scale, red being the point of highest potential in the system while blue is the lowest one.  \textbf{c,} Output characteristics of the Y-shaped device pictured in a). The curve shows an asymmetric electrical behavior related to the distribution of ions in the system, as illustrated in the subset figures. When the applied voltage is negative, the output current reaches a plateau at about 0.4 V, whereas it remains linear when V > 0 and up to 0.9 V. The rectification coefficient, computed as $|I_{OUT}|_{0.9V}/ |I_{OUT}|_{-0.9V}$, is equal to 4.4. \textbf{d,} Breakdown of the rectification coefficients as a function of which terminals are polarized. The x-axis indicates which electrodes are addressed in each configuration (e.g.,(1·2)-3). The grounded electrode is always written to the left of the hyphen (electrodes 1 and 2 are shorted), and the terminal undergoing  voltage sweep is always written to the right (electrode 3). The numbers correspond to the electrodes indicated in the inset of the figure.}
  \label{fig:fig1}
\end{figure}

Within a network of physically connected polymer fibers immersed in an aqueous electrolyte, the distribution of potentials plays a key role, influencing the behavior of the system and the cross-talking between the different dendrites of the network. Figure~\ref{fig:fig1} presents the nonlinear electrical behavior of a Y-shaped dendritic system grown between four gold electrodes of a typical MultiElectrode Array (MEA), operated within a water-based electrolyte (Phosphate Buffered Saline, PBS). Figure~\ref{fig:fig1}b depicts how one can influence the way the system behaves by carefully choosing which electrodes to address. Indeed, depending on the polarization of the nodes of the circuit, the electrochemical doping profile (supported by the cationic charges in water) within the different PEDOT:PSS branches can be modified. This allows the control of different conductance profiles within a single Y-shaped dendritic topology inducing various transport paths in the material, which is made possible only because the object has a shape of its own. As such, a polymer dendrite cannot be considered as a discrete element with a well-defined number of inputs and outputs, since a direct structure-properties relationship emerges from these kinds of devices.\\
 
Figure~\ref{fig:fig1}c shows the asymmetric behavior of the Y-shaped network when pinning the central electrode of the dendritic system to the ground while varying a single input between -0.9 V and 0.9 V. When the input voltage is negative, cations close the channel between the input and the dendrite core. As a single input channel has a residual volume compared to the full dendritic system, the total capacitance of the dendrites that do not participate in the conduction of the current is superior to that of the channel separating the single input from the dendrite core. This results in a self-gating effect on a single dendritic extension, with cations accumulating in the conduction path and the current starting plateauing around -0.4 V. On the other end, when the input potential V is positive, thanks to the distribution of capacitance within the system, cations are distributed among the three dendrites of the multiterminal OECT. Consequently, the dedoping of the conduction dendrite is far less important than observed when V < 0 V, and its conductance remains stable up to 0.9 V.\\

This Y-shaped system thus presents a forward and a reverse direction, similar to a diode. The rectification coefficient, defined here as $|I_{OUT}|_{0.9V}/ |I_{OUT}|_{-0.9V}$, is presented for all the permutations of the applied potential and ground in the system. Figure 1d presents the different ratios obtained in these different configurations. The higher rectification is obtained for the previously discussed case, with a ratio of 4.4, while in one of the other cases, when the lower right electrode is grounded and V is applied to the left electrode, a symmetrical (although nonlinear) behavior was observed, resulting in a rectification coefficient equal to one. These differences further highlight the intricate relationship between topology and nonlinearity: the specific location of the input conditions the output of the system.

\subsection{Inter-gating effect}
If several dendrites physically sharing the same node are intricated, their electrochemical interconnection with the common electrolyte is a frequent issue provoking cross-talking. Because of the common electrolyte, one of the most critical technological challenges for OECTs is the ability to independently address each device of the matrix, which is required to access a single element’s information. This issue can be addressed at the expense of more complex fabrication procedures\cite{spyropoulos_internal_2019, jeong_ion_2023}.\\

\begin{figure}[h]
	\centering
	\includegraphics[width=1\columnwidth]{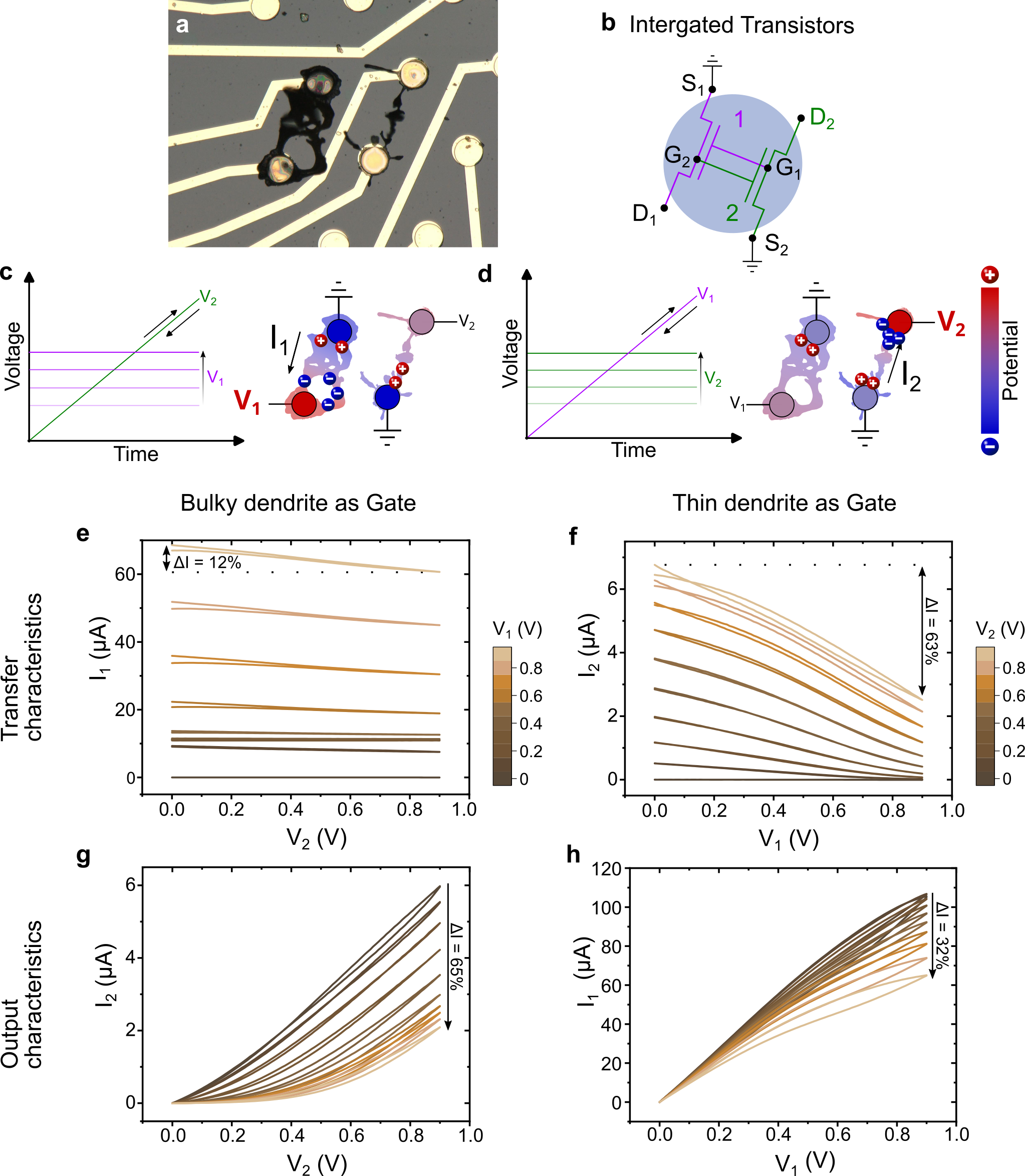}
	\caption{\textbf{Inter-gating effect between two parallel dendritic OECTs.} The morphology of the PEDOT:PSS fibers is of primary importance. \textbf{a,} Microscopic photo of the two dendrites used as OECTs, showing that the left one (grown at 25 Hz) is noticeably thicker than the right one (grown at 200 Hz). \textbf{b,} Equivalent circuit of the two parallel transistors. The PEDOT:PSS channel of each OECT act has the gate from the other transistor. \textbf{c-d,} Ionic distribution when the bulky and thin dendrites are respectively used as the gate. The color scale indicates the nodes of highest potential (red) and lowest potential (blue) in the circuit. \textbf{e-f,} Transfer characteristics of the thin and bulky dendrites, respectively. \textbf{g-h,} Output characteristics of the thin and bulky dendrites, respectively.}
	\label{fig:fig2}
\end{figure}

On the other hand, cross-talking between devices can also be apprehended as communication mean between physically disconnected parts of the system. PEDOT:PSS being a good material for efficient OECT gating\cite{koutsouras_efficient_2021, tarabella_effect_2010}, two dendritic transistors operating simultaneously should therefore have a gating effect on each other, as displayed in Figure~\ref{fig:fig2}. 

	The influence of a dendritic device over another was assessed by growing two parallel dendrites, with the first being noticeably thicker than the second (Figure~\ref{fig:fig2}a).\\
	 
	The dendrites were operated simultaneously and characterized as OECTs, with their transfer and output characteristics recorded (Figure~\ref{fig:fig2}e-h). Once again, the morphology of the dendrites appears to play a major role in the way the whole system behaves. Indeed, using the thicker dendrite as the gate rather than the thinner one ($V_1$ fixed while $V_2$ was swept, see Figure~\ref{fig:fig2}c) has a greater impact on the output current. Figure~\ref{fig:fig2}g shows that as $V_1$ increases, the drain current $I_2$ gradually decreases. The same trend is also observed when the roles are reversed and the thinner dendrite is used as the gate (see Figure~\ref{fig:fig2}h), although to a lesser degree: in the former case, the drain current decreases by 65\% when $V_1$ = 0.9 V versus when $V_1$ = 0 V, compared to around 32\% in the latter case (Figure~\ref{fig:fig2}h). This effect is not due to the fatigue of the material (Figure~\ref{fig:figS1}).\\
	
	The shape of the output characteristics is also distinctly different. When the bulky dendrite is used as the gate, $I_2$ remains very low up to around 0.4 V, before entering a linear region. Interestingly, this was not observed in the second case (Figure~\ref{fig:fig2}h), during which the output current follows a linear trend before starting to saturate, although the saturation region is not reached yet at 0.9 V. Once again, the morphology of the dendrites and their ability to retain ions might be at the origin of this phenomenon. With its large volumetric capacitance, PEDOT:PSS is an effective gate material that can efficiently promote doping/dedoping. The voluminous fiber is thereby able to dedope its neighboring dendrite almost entirely at high enough $V_1$, as the lowest potential of the system is in that case the grounded electrode of the thin dendrite (Figure~\ref{fig:fig2}c). $V_2$ has to overcome a certain threshold to be able to move ions back into the system and to start regenerating the initial high doping state of the dendrite. When the roles are reversed (Figure~\ref{fig:fig2}d), the thin dendrite also plays the role of the gate of the system, although less efficiently than its more voluminous counterpart as it is not able to close entirely the channel of the second transistor.\\
	 
	The previous sections have established that a network of conductive polymer dendrites behaves as a multiterminal transistor, within which the dendrites are not a collection of independent elementary nodes, but are rather interconnected and interdependent objects. If a system made of multiple dendrites physically connected experiences a self-gating effect, two independent devices sharing the same electrolyte also undergo a similar inter-gating effect. As a consequence of this cross-talking, computing can be implemented on systems of electrically interconnected dendrites (forming a single continuous topology) as well as systems of distinctive dendritic topologies electrochemically connected with an electrolyte.

\subsection{Multiply-accumulate operation with multiterminal OECTs}
\begin{figure}[h]
	\centering
	\includegraphics[width=1\columnwidth]{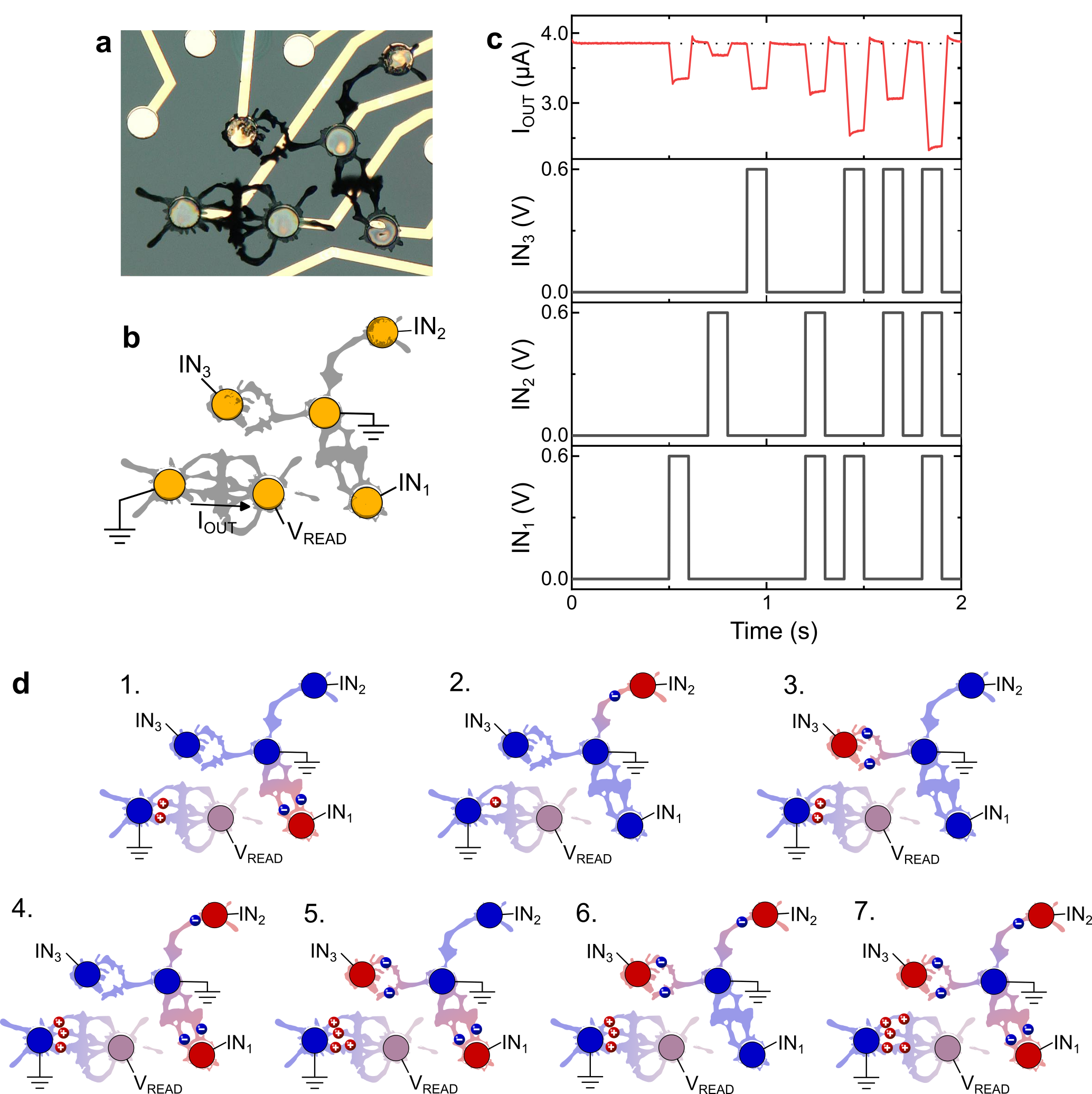}
	\caption{\textbf{Real-time computing, taking advantage of the non-linearities and inter-gating effect of the dendritic OECTs. a-b,} Microscopic photo and schematics of the system showing the three inputs and the output. The Y-shaped device is used as the input, while the output current is read at the drain of the single dendrite ($V_{READ}$ = 100 mV).  \textbf{c,} Output current and input voltages as a function of time. Transients were manually removed to focus on the steady-state response of the circuit. \textbf{d,} Ionic distribution schematizing the dedoping mechanisms in the system when the inputs are polarized. The color scale indicates the nodes of highest potential (red) and lowest potential (blue) in the circuit.}
	\label{fig:fig3}
\end{figure}

Taking advantage of the inherent properties of dendritic networks, a multiterminal system was used to implement the multiply-accumulate (MAC) function (Figure~\ref{fig:fig3}).  Monitoring only one output, it becomes possible to detect what combination of inputs was activated. The circuit was made of a Y-shaped device as the inputs, and the output current was read from the drain electrode of a single dendrite (Figure~\ref{fig:fig3}a-b), all immersed in PBS. Pulses of voltage (0.6 V, 200 ms) were sent as inputs to the outermost electrodes of the Y-shaped device, the central terminal being grounded, in the fashion depicted in Figure~\ref{fig:fig3}c, while a small potential was applied to the output electrode ($V_{READ}$  = 100 mV). 
As a result of the positive voltage pulses, the output dendrite experienced electrochemical dedoping due to the movement of ions through the electrolyte, as schematized in Figure~\ref{fig:fig3}d-j. The modulation of current is characteristic of the activated input node: the output current undergoes a 13\% modulation for IN1, 5\% for IN2 and 18\% for IN3. The magnitude of the modulation appears to be related to the morphology of the dendrites, as seen in Figure~\ref{fig:fig2}, as well as the distance that separates the input node from the output dendrite, the third input being the closest one to the output dendrite while IN2 is the farthest away. This dependency of the gating effect with distance from the drain electrode had already been observed in thin-film OECTs, and proved useful for orientation selectivity\cite{gkoupidenis_orientation_2016}.

Here, as seen in Figure~\ref{fig:fig3}c, every single combination of inputs has a unique signature that can be sensed by the output dendrite. The more inputs are activated, the more important is the modulation of the output current. When all three inputs are set to 0.6 V, this modulation reaches a maximum of 38\%. Therefore, with a read-out from a single node in the circuit, the system is able to discriminate which inputs have been activated. Such operation reproduces the multiply-accumulate (MAC) function that is a key operation in artificial neural networks. In this particular implementation, the weight is defined by the topology of the dendrites belonging to the Y-shaped device, while the accumulation is implemented with the read-out device. Nevertheless, it is important to note that, as for now, the stochasticity involved in the dendritic growth process implies that controlling with precision the weight of each dendrite is not so straightforward. As it has been demonstrated, the morphology of dendritic objects can be controlled globally. However, at the local scale, the stochastic nature of dendritic growth intrinsically promotes large variabilities, conflicting with the idea of physical reproducibility (and a fortiori of the reproducibility of electrical properties). Moreover, when operating with a higher number of polarized nodes in the system, growing a perfectly predefined network of dendritic OECTs can become delicate. From a general perspective, MAC operation corresponds to a linear function that could be rationalized for a small topological dendritic system (i.e., three branches of the Y-shaped device). In the following of the discussion, we show that more complex topologies can lead to nonlinear functions that increase the computational power of the dendritic network. 

\subsection{\textit{In materio} computing}
The influence of the system on itself and on the nearby PEDOT:PSS dendrites, as well as its ability to retain information over time and discriminate past voltage events, are essential ingredients for the realization of complex information processing tasks. Indeed, performing time series prediction thanks to the nonlinear dynamics of such an organic electrochemical network\cite{cucchi_reservoir_2021}, coupled with the structural plasticity offered by this type of hardware\cite{janzakova_structural_2023}, makes dendritic networks an exciting candidate to instill a new direction in the search of neuromorphic material for the hardware revolution that is eagerly waited for. In this sense, conductive polymer networks could find themselves at the crossroad between different domains of research, borrowing ideas both from the biological realm (neurogenesis and synaptic plasticity) and computing. Yet so far, if dendritic networks appear to be a very useful tool to project information in a reservoir of higher dimension, a subsequent layer of machine learning is still required to perform classification\cite{cucchi_reservoir_2021}. In this work, we propose a new framework based solely on hardware, capable of in memory spatiotemporal information processing, with the output condensing in a single value of current. 

\subsubsection{Spatial information processing}
First, we studied the response of a dendritic network to a single sequence of inputs, projected on different input electrodes. This sequence consists of eight patterns of three bits, always presented in the same chronological order, as introduced in Figure~\ref{fig:fig4}. The bits are encoded as a potential vs ground, negative to represent ‘0’ and positive to represent ‘1’. Each voltage pattern is applied during ten seconds (“WRITE” operation) and followed by a brief “READ” operation of about 50 ms, during which the output dendrite is biased ($V_{DS}$=100 mV). Between two ‘WRITE’ operations, the system is at “rest” by setting the three input electrodes and the two terminals of the output dendrite to the ground for ten seconds.\\

As the input voltage sequence is repeated, a current modulation pattern appears in the output of the system. This cycle is best observed in the normalized currents (Figure~\ref{fig:fig4}), where the unique signature can be appreciated, and the drift observed in the output current has been removed (see Figure~\ref{fig:figS2}). The modulation of current is computed as follows:

\[\frac{\Delta I}{I}=\frac{I_{READ} - I_{REST}}{I_{REST}}\]
  
When the same temporal sequence is projected on different input electrodes, the output of the network is modified. As this signature is related to the ability of the network to distribute ions between the different dendrites, and thus to the morphology of the network, this suggests that no two electropolymerized systems will give the exact same response to similar inputs, thanks to the stochasticity observed in the electropolymerization process\cite{ciccone_growth_2022, cucchi_directed_2021}. The possibility to build this type of hardware on demand could have very powerful implications for security purposes, notably in the quest for physically unclonable functions.\\

Interestingly, even when two spatial projections are quite similar, the signature of the system in the output current can be told apart: for spatial projections \#1 and \#3 (SP1 and SP3), only one of the input electrodes differs (Figure~\ref{fig:fig4}e), yet in both cases the output current variations are clearly distinguishable. Figure~\ref{fig:fig4}d indeed shows that each spatial projection signature is different and consistently reproducible. 
In addition, this variation depends on the position of the input electrodes. The maximum increase in current is reached at step 3 for SP1 (about 10\%), whereas it happens at step 6 for SP2 and SP3 and is more consequent (slightly superior to 15\%). In other words, the same voltage pattern does not have the same effect on the output of the system depending on which input electrodes are addressed. This means that, given only the output current on one electrode, the system would be able to discriminate between different sources of information. This amounts to a wireless electrochemical communication system, where information can be transmitted though the electrolyte, and each source of information could be identified by its unique signature. Such spatial projection of the input patterns also demonstrates that nonlinear functions emerge when the topology becomes more complex. In comparison to the linear MAC operation demonstrated with the Y-shaped devices, multi-terminal devices cannot be rationalized as a linear combination of the input. In fact, the (1,1,1) and (0,0,0) patterns do not produce the maximum and minimum modulation of current as expected from a linear combination of input patterns.

\begin{center}
	\includegraphics[width=0.8\columnwidth]{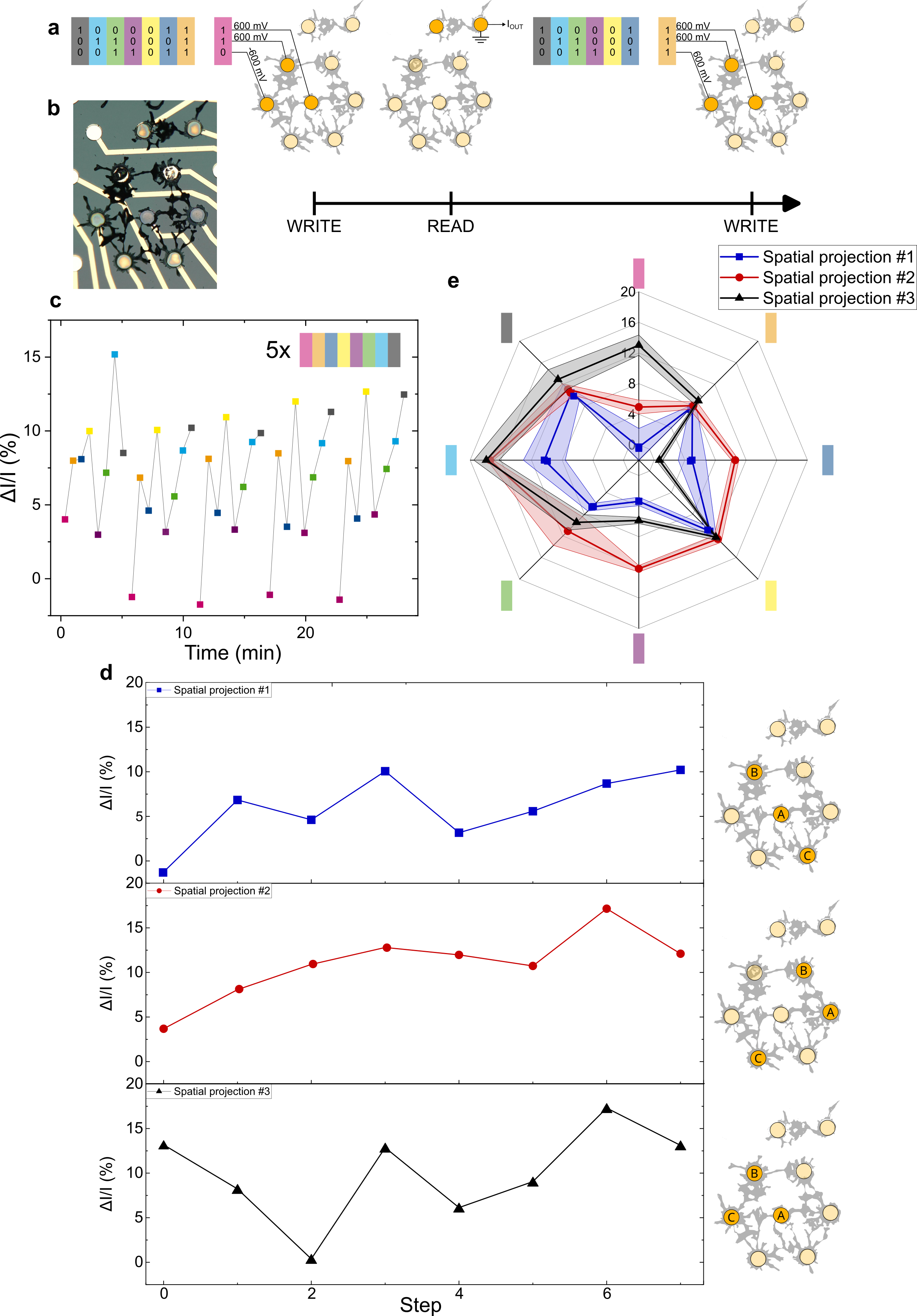}
	\captionof{figure}{\textbf{In memory spatial information processing. a,} Principle diagram showing the WRITE and READ operation sequence. Each color stands for a 3-bit voltage pattern. \textbf{b,} Microscopic photo of the dendritic network used for spatial information processing. \textbf{c,} Output current variation for five repetitions of Spatial Projection \#1. The output current variation is computed as the difference between the output current measured after a WRITE operation and the previous REST operation (all terminals set to the ground for ten seconds). \textbf{d,} Output current variation for different sets of inputs receiving the temporal sequence presented in a). In the schematics of the dendritic system on the right, each letter represents the position of a single bit in the 3-bit pattern (A,B,C). \textbf{e,} Spider diagram showing the average current variation and standard deviation over five cycles for three different sets of inputs, as presented in d). Chronological order is shown clockwise, starting from the top of the diagram.}
	\label{fig:fig4}
\end{center}

\subsubsection{Temporal information processing}
\begin{figure}[h]
	\centering
	\includegraphics[width=1\columnwidth]{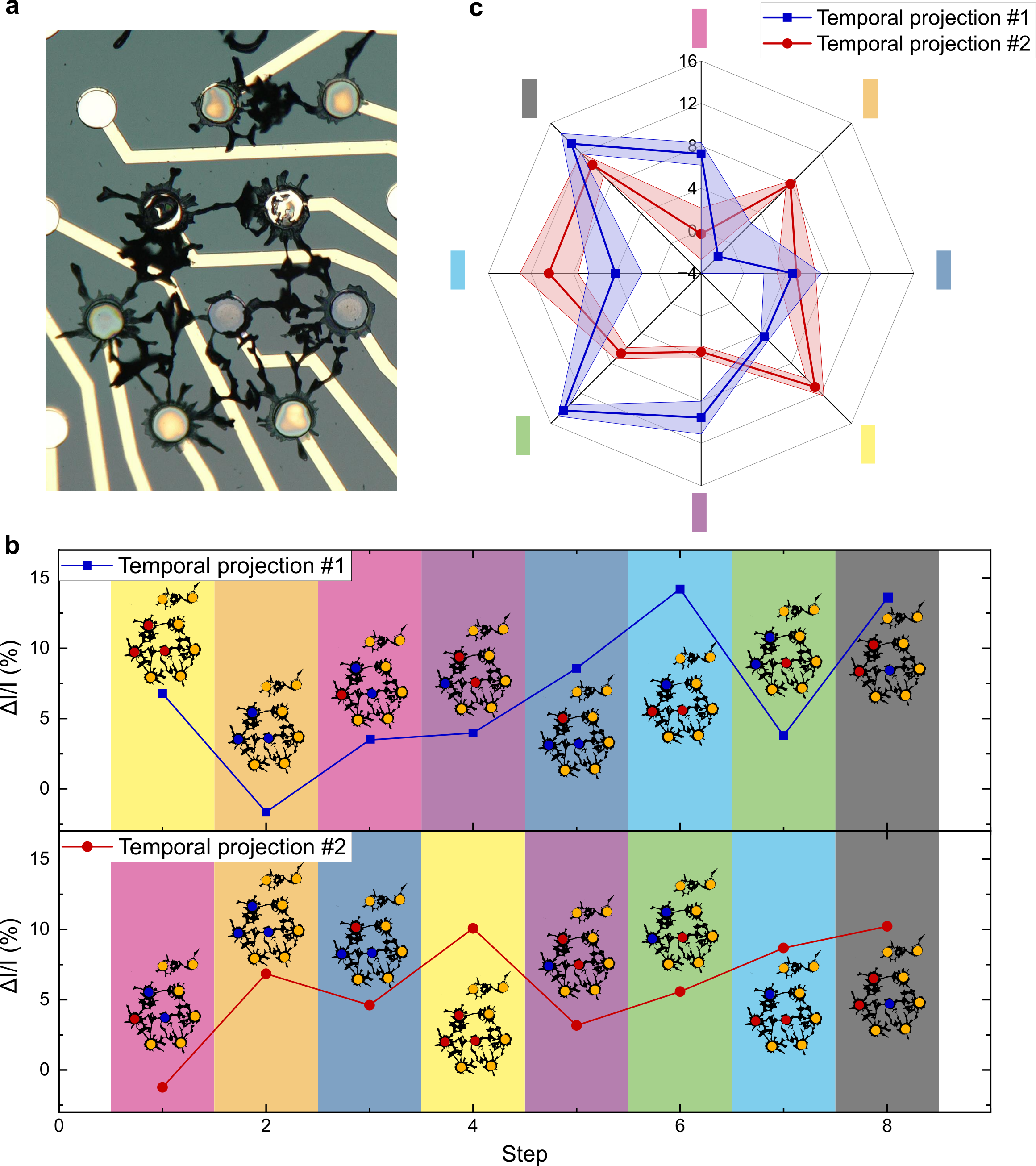}
	\caption{\textbf{In memory temporal information processing.  a,} Microscopic photo of the dendritic network used for temporal information processing, the same as in Figure 4. \textbf{b,} Output current variation for different temporal sequences (the 3-bit patterns ore the same, only shuffled) presented to the same input electrodes in the system. Each color represent a 3-bit pattern. Inset : schematic representation of the dendritic system, for which blue represents a HIGH logic state (physically encoded as 600 mV), and red represents a LOW logic state (physically encoded as -600 mV). \textbf{c,} Spider diagram showing the average current variation and standard deviation over five cycles for two different temporal sequences. Note that following the diagram clockwise does represent the chronological order for temporal projection \#2, but not for \#3.}
	\label{fig:fig5}
\end{figure}

If spatial information discrimination allows for the unique identification of the emitting source, the dendritic network is also able to encode temporal information. Indeed, Figure~\ref{fig:fig5} shows that, even if two sequences contain the same input nodes, the chronological order in which they are presented to the system may influence the output. Using the same structure as for spatial information processing, we demonstrate here that the output of the system is state dependent. 
Figure~\ref{fig:fig5}b shows that a single voltage pattern can have very different contributions to the output of the system based on the history of the inputs that have previously been presented to the structure. For example, for both temporal projections, the second voltage pattern presented to the dendritic network is identical (the (1,1,1) pattern represented by the orange color). In the first case, the system appears to have been potentiated by the first pattern of the sequence. Therefore, the second voltage pattern has a negative contribution to the output current. However, for the second temporal projection, the output dendrite has already been depressed by the first voltage pattern of the sequence. As a consequence, the modulation of current following second pattern of the sequence (the orange one) is in that case positive.  
Each voltage pattern thereby has a different influence on the output current variation based on the internal state of the system. Figure~\ref{fig:fig5}c shows that, in order to be interpreted correctly, the whole input sequence needs to be captured. Indeed, reading the output related to only one voltage pattern will not allow to identify that voltage pattern, as its signature may vary depending on the internal electrochemical doping state of the output dendrite, and thus on the previous voltage events that took place within the system. If the network happened to be memoryless, the output current variation would no depend on the chronological order of the input sequence, as the variation of current is always calculated compared to the previous resting state, as explained in the previous section. 

This behavior of dendritic networks could have a tremendous impact on information processing: by only reading the output current of a single dendrite, within a characterized dendritic network, a dendritic device would be able to provide both spatial and temporal information. This is due to the highly complex and interwoven interactions that arise within such networks sharing a common electrolyte, a communication medium that allows the exchange of information between physically disconnected parts of the network.

\subsection{Hardware-dependent functions}
Exploiting the intrinsic randomness of a manufacturing process to generate a unique output considered to be a signature of a device is an idea that has important repercussions for security purposes, and that has already been explored to implement Physical Unclonable Functions\cite{bohm_physical_2013, gao_physical_2020}. In that regard, electropolymerized networks have plenty to offer, given the hardware-dependent functions that emerge as a consequence of the particular distribution and morphology of the polymer fibers within the network. The patterns emerging from the spatial and temporal projections in the previous section of this article could indeed be considered as a signature, a unique identification key: in order for the desired output to be generated, the correct input has to be sent to the network, both in terms of the combination of input electrodes but also of the temporal sequence. Moreover, as electropolymerization is an easy-to-perform process, this means that the signature of the device can be generated by the end-user itself, making him the only one in possession of the unique signature of the device.\\

\begin{figure}[h]
	\centering
	\includegraphics[width=1\columnwidth]{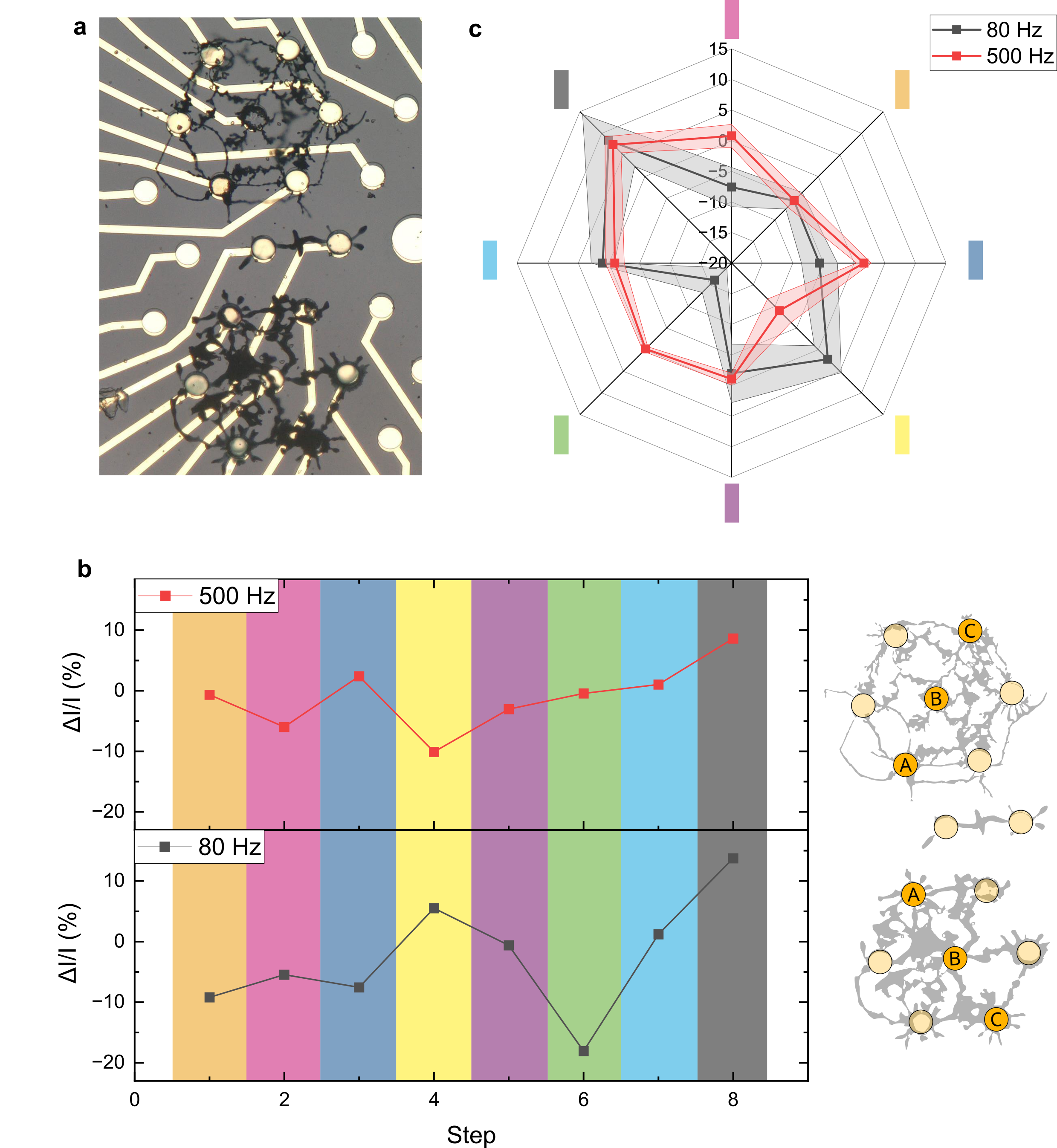}
	\caption{\textbf{Influence of the morphology of the dendrites on the electrical behavior of the system. a,} Microscopic picture of the two dendritic systems, grown at 500 Hz (top) and 80 Hz (bottom). The readout is the single dendrite in the middle, grown at 80 Hz. \textbf{b,} Output current variation of the two devices over one cycle, for the same input sequence projected. The sets of electrodes were chosen to be symmetrical relative to the readout dendrite. \textbf{c,} Spider diagram showing the average and standard deviation of the electrical behavior of the two devices computed over 8 cycles. Chronological order is shown clockwise, starting from the top of the diagram. }
	\label{fig:fig6}
\end{figure}

In this section, we demonstrate that dendritic networks with similar topologies still exhibit distinct functional properties. Two different networks were grown, one with thick dendrites (growth frequency of 80 Hz) and the other with thinner objects (growth frequency of 500 Hz). Both networks were grown to have a comparable topology, on each side of a unique dendrite that was used as the readout of both systems (see Figure~\ref{fig:fig6}a).\\

The same sequence of patterns as presented in Figure~\ref{fig:fig4} was applied to both networks sequentially. This resulted in two different outputs, based on which network was addressed, as shown in Figure~\ref{fig:fig6}b and Figure~\ref{fig:fig6}c. While the signatures of some of the patterns were very similar for both networks (notably the orange, purple, light blue and grey ones), the behavior was divergent for the other inputs. Indeed, the green and pink patterns triggered a positive modulation of the output current for the 500 Hz network, while in the 80 Hz network they had a negative contribution. Consequently, over the whole sequence, each network possesses its own unique signature, as shown in Figure~\ref{fig:fig6}c. The two networks, although they share the same electrolyte and the same readout dendrite, still can be differentiated by their electrical behavior even when presented identical inputs. Thanks to this property, a single readout device could thus be used to read a whole constellation of dendritic networks sharing the same medium, as each network would have a unique input-output relationship with the readout.\\
 
Two important conclusions can be drawn. First, although the morphology of the fibers themselves is impossible to control finely, as it has been mentioned elsewhere\cite{janzakova_analog_2021}, it is still possible to define the topology of the network from a global standpoint. Indeed, the connections between the nodes has essentially been maintained: although the conductance of a specific dendrite might be harder to define, two nodes can still be connected from a binary perspective. Second, two different networks, however similar they might look, are not identical. If this holds from a physical point of view, this is even clearer when comparing their electric behavior. This ensures the unicity of every network grown on such a platform as the one presented in this article, which is at the same time a boon (e.g., for cryptographic functionalities or classification problems) and a bane for applications such as deterministic logical functions. 

\subsection{Discussion}
Biological neural networks are an important source of inspiration when it comes to efficient computing, mainly due to their innate ability to massively parallelize information processing. If large artificial neural network models have proven to be very efficient computationally, they are still far from the energy efficiency of biological brains. In these models, performance comes with high accuracy in the synaptic weights and dense interconnections between layers of neurons that translate into important energy consumption. In biological networks, only useful connections are grown and result in sparse and specialized topologies that optimize resources utilization. Additionally, these topologies possibly bear more profound memory representation that connectionist models based solely on synaptic weight evolution during learning cannot capture. Topologies set the dimensional space in which synaptic plasticity can be expressed, and could possibly be the foundation for semantic representations (i.e., higher level of organization in the representation of concepts)\cite{von_der_malsburg_toward_2021}. Our work highlights the role of topologies in the spatiotemporal projection of information and could be a framework to reconciliate connectionist and semantic approaches. We show in this paper that the control operated over the topology of dendritic networks governs the complexity of the operation of the system, due to the interplay between form and function. In other words, if small networks implement nearly linear functions, such as the MAC operation discussed earlier, more developed systems exhibit nonlinear behaviors that are more complex to implement but offer greater computational power. The bottom-up approach of dendritic growth therefore allows the function of a multiterminal device to be defined by its physical structure, as complexity emerges from the physical structure of the network itself.\\
 
Another concept seldom discussed in the computing community is the wide diversity of synapses and information carriers in the brain. Although a great number of neuromodulators has already been identified\cite{cuevas_neurotransmitters_2019}, our understanding of the underlying mechanisms by which they participate in information processing remains unclear. Especially striking is the varied nature of these, from peptides to gaseous molecules. Apart from neurotransmitters, information also appears to be conveyed by marker molecules, as it has been proposed that their gradient of concentration could guide axonal growth in the early stage of brain development\cite{stoeckli_understanding_2018}. This sharply contrasts with the current view in computing, where electrons are the only information carrier. Using an electrochemical platform such as the one presented in this article, the number of information carriers in the system could be increased drastically, as many biomolecules have already been demonstrated to modulate the behavior of organic electrochemical transistors\cite{gualandi_selective_2016, burtscher_sensing_2021, keene_biohybrid_2020, xie_organic_2020, strakosas_organic_2015}. With recent advances regarding degradability and disintegration of organic polymers\cite{rai_future_2023, teo_towards_2023, tropp_design_2021}, full structural plasticity could also become a way to store information directly in the structure of the network, making the most out of the limited resources of the system by destroying unnecessary synapses to reemploy the monomer molecules more efficiently.\\

The ability to learn and evolve is indeed an incredibly potent property of the brain, one that the current top-down approach of solid-state electronics cannot emulate. Evolving systems have the great advantage to adapt to the tasks at hand, so that resources are employed as efficiently as possible. In such systems, complex behavior arises from simple local rules, as it was proposed with cellular automata\cite{wang_elementary_2024}. Using the computational power of adaptive bottom-up physical systems has been explored with NanoCells\cite{skoldberg_nanocell_2007, tour_nanocell_2003, tour_nanocell_2002} or slime mold\cite{adamatzky_physarum_2007, adamatzky_programmable_2010, adamatzky_road_2010, reid_slime_2012, nakagaki_smart_2001}, but the physical structures in these examples were obtained without any physical relationship with the computation to be performed. Here, the versatility of electropolymerization\cite{janzakova_analog_2021} offers to adjust the operation of the device as needed in operando, enhancing its capabilities and bringing organic electronics one step closer to biology. Implementing this technology in a solid-state electrolyte could make solid-state electronics more plastic and create new possibilities with the integration of a bottom-up additive approach to otherwise deterministic electronics. Indeed, the stochastic growth and the variability of organic electronics, which are inconveniences in the realm of traditional electronics and digital computing, turn into opportunities when it comes to unconventional computing. Having systems which growth is not fully controllable, and which behavior cannot be anticipated is highly valued for cryptographic and security applications such as Physical Unclonable Functions (PUFs), which take advantage of the variability in production to generate devices that present a unique signature\cite{bohm_physical_2013}. Although still a recent concept, conductive polymer dendrites show great potential for the physical implementation of unconventional computing and the integration with solid-state electronics. 

\section{Conclusion}
Our work establishes electropolymerized PEDOT:PSS dendritic networks as a new class of morphology-dependent hardware, which properties not only arise from the material but also from the morphology of the fibers and the topology of the network. We highlight the complex interplay between form and function in these structures.\\

We bring to light that these objects present a nonlinear and asymmetric electric behavior directly related to their morphology and the ionic distribution across the dendrites, which depends on the activity of all the fibers of the network. PEDOT:PSS dendritic networks also appear to possess an intrinsic memory that, coupled with the self-gating and inter-gating effects of the polymer dendrites, allows the system to perform \textit{in materio} computing tasks, such as spatiotemporal information processing. Indeed, conductive polymer networks are able discriminate the source of surrounding voltage events and present state-dependent computing abilities, as in such systems the output of the network seems to be conditioned by past inputs. Finally, we evidence that these properties, coupled with the stochasticity of AC-electropolymerization, allows the growth of networks which topologies might be very similar on a large scale, but in which the distinct morphology of each dendrite gives rise to a unique granularity, and in consequence a distinctive electric behavior.\\

Electropolymerization definitely emerges as a fabrication method of choice in the quest for a new paradigm that could help solve the hardware crisis at hand, as it offers the possibility to create complex structures with minimal resource requirements, implementing an effective bottom-up approach. With close to no need for complex and costly machines and incredible simplicity, it has the potential to foster a new era of organic, cost-effective and bioinspired electronics.

\section{Materials and methods}

\subsection{Substrate fabrication}
The MEAs used as a substrate for the electropolymerization of dendritic networks were fabricated as described elsewhere\cite{ghazal_electropolymerization_2023, ghazal_precision_2023}.

\subsection{Dendritic growth}
Dendritic growth was carried out through AC-electropolymerization using a Keysight 33600A Trueform waveform generator. Square waves with a frequency of 80 Hz and a peak voltage amplitude of 5 Vp were the default waveform applied between the working and grounded electrodes, unless otherwise specified. For the growth to happen, the system was immersed in an aqueous solution containing 1 mM of poly(sodium-4-styrene sulfonate) (NaPSS), 10 mM of 3,4-ethylenedioxythiophene (EDOT), and 10 mM of 1,4-benzoquinone (BQ). All chemicals were bought from Sigma Aldrich and used without further modification. The growth of the system was monitored though a TrueChrome Metrics camera.

\subsection{Electrical characterization}
All the electrical characterizations presented in this work were conducted on an Agilent B1500A Semiconductor Analyzer coupled with a B2201A Switching Matrix.

\subsection{Microscopic images}
All microscopic pictures were taken using a Keyence VHX-6000 microscope.

\section*{Acknowledgments}
The authors thank the RENATECH network and the engineers from IEMN for their support. This work is funded by ERC-CoG IONOS project \#773228., and by the Région Hauts-de-France. FA thanks the MEIE from Quebec for support of the neuromorphic computing chair. CS thanks Anne-Sophie Vaillard for her precious help with the fabrication of the MEAs.

\printbibliography

\newpage
\section*{Supplementary Materials: Brain-inspired polymer dendrite networks for morphology-dependent computing hardware}
\beginsupplement

\begin{figure}[h]
  \includegraphics[width=\columnwidth]{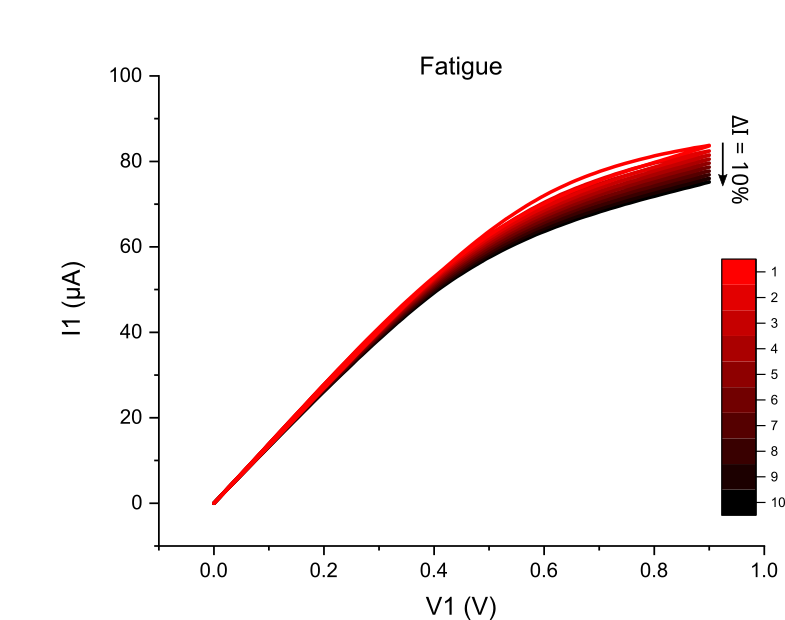}
  \caption{\textbf{Fatigue effect on the more voluminous PEDOT:PSS dendrite} presented in Figure~\ref{fig:fig2}. At 0.9 V, the drop of current represents 10\% of the initial value.}
  \label{fig:figS1}
\end{figure}

\begin{figure}[h]
  \includegraphics[width=\columnwidth]{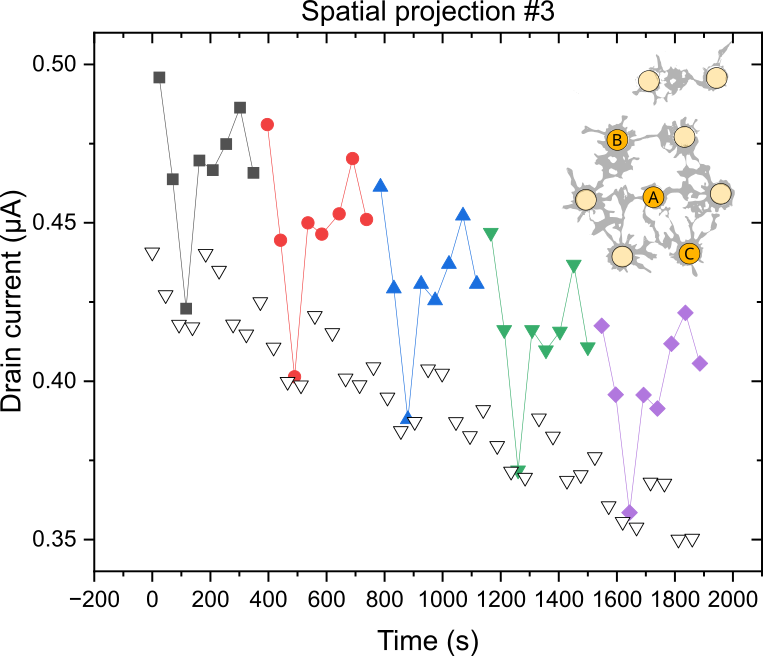}
  \caption{\textbf{Drain current of the output dendrite of the dendritic system used for the spatiotemporal information processing.} This graph brings to light the pattern that appears in the output current when an input sequence is repeated. The downward current drift discussed in the text is clearly observable. Here, each color represents a repetition of the input sequence, namely Spatial Projection \#3.  The white down-pointing triangles represent the output current after a ‘REST’ operation, which are used to compute the output current variations ($\Delta I/I$).}
  \label{fig:figS2}
\end{figure}

\begin{figure}[h]
	\includegraphics[width=\columnwidth]{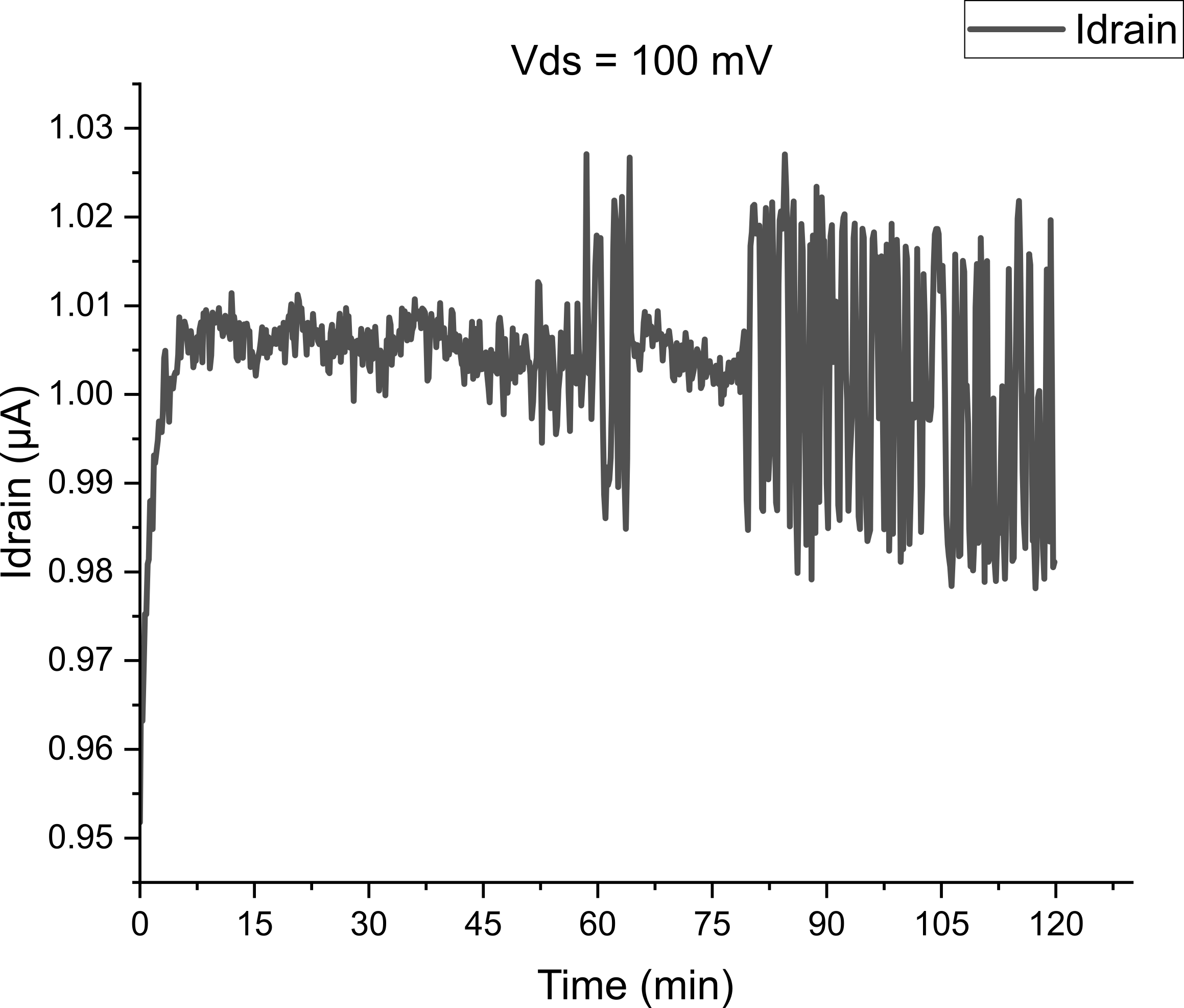}
	\caption{\textbf{Aging test in PBS}. A bias of 100 mV was applied across the terminals of the readout dendrite used in Figure~\ref{fig:fig6}, and the drain current was recorded. Contrary to what was observed during the \textit{in materio} programming experiments, no consequent drift appeared during the 2h recording, suggesting that the drift observed in Figure~\ref{fig:figS2} is not due to the aging of the material.}
	\label{fig:figS3}
\end{figure}

\subsection*{Origin of the drift}
Since it does not appear to be due to the ageing of the material (Figure~\ref{fig:figS3}), this slow decay observed in the output current might most probably come from an imbalance between the doping and dedoping mechanisms induced by the applied programming sequence. As suggested earlier, all cations entering the material might not drift away easily, therefore keeping the material in an increasingly undoped state, which would also explain the fatigue observed in Figure~\ref{fig:figS1}.

\end{document}